\documentclass[aps,prx,groupedaddress,twocolumn,superscriptaddress]{revtex4-1}
\usepackage{graphicx,multirow}
\usepackage{amsmath}
\usepackage{color,soul,array}

\newcommand{\TN}{$T_{\mathrm{N}}$}
\newcommand{\Gfour}{$\Gamma^{\mathrm{4g}}$}
\newcommand{\Gfive}{$\Gamma^{\mathrm{5g}}$}

\newcommand{\sxy}{$\sigma_{xy}$}
\newcommand{\sxx}{$\sigma_{xx}$}
\newcommand{\sxxzero}{$\sigma_{xx,0}$}
\newcommand{\sxyint}{$\sigma_{xy}^{\mathrm{int}}$}
\newcommand{\pxy}{$\rho_{xy}$}
\newcommand{\pxx}{$\rho_{xx}$}

\begin{document}

\title{The anomalous Hall effect in non-collinear antiferromagnetic Mn$_{3}$NiN thin films}

\author{David Boldrin}
\email[Corresponding author: d.boldrin@imperial.ac.uk]{}
\affiliation{Department of Physics, Blackett Laboratory, Imperial College London, London, SW7 2AZ, UK}
\author{Ilias Samathrakis}
\affiliation{Institute of Materials Science, TU Darmstadt, 64287, Darmstadt, Germany}
\author{Jan Zemen}
\affiliation{Faculty of Electrical Engineering, Czech Technical University in Prague, Technick\'{a} 2, Prague 166 27, Czech Republic}
\author{Andrei Mihai}
\affiliation{Department of Materials, Imperial College London, London, SW7 2AZ, UK}
\author{Bin Zou}
\affiliation{Department of Materials, Imperial College London, London, SW7 2AZ, UK}
\author{Bryan Esser}
\affiliation{Center for Electron Microscopy and Analysis, 1305 Kinnear Road, Columbus, OH 43212, United States of America}
\author{Dave McComb}
\affiliation{Center for Electron Microscopy and Analysis, 1305 Kinnear Road, Columbus, OH 43212, United States of America}
\author{Peter Petrov}
\affiliation{Department of Materials, Imperial College London, London, SW7 2AZ, UK}
\author{Hongbin Zhang}
\affiliation{Institute of Materials Science, TU Darmstadt, 64287, Darmstadt, Germany}
\author{Lesley F. Cohen}
\affiliation{Department of Physics, Blackett Laboratory, Imperial College London, London, SW7 2AZ, UK}

\date{\today}

\begin{abstract}
We have studied the anomalous Hall effect (AHE) in strained thin films of the frustrated antiferromagnet Mn$_{3}$NiN. The AHE does not follow the conventional relationships with magnetization or longitudinal conductivity and is enhanced relative to that expected from the magnetization in the antiferromagnetic state below $T_{\mathrm{N}} = 260$\,K. This enhancement is consistent with origins from the non-collinear antiferromagnetic structure, as the latter is closely related to that found in Mn$_{3}$Ir and Mn$_{3}$Pt where a large AHE is induced by the Berry curvature. As the Berry phase induced AHE should scale with spin-orbit coupling, yet larger AHE may be found in other members of the chemically flexible Mn$_{3}A$N structure. 
\end{abstract}

\maketitle

The realization of large anomalous Hall effects (AHE) in antiferromagnets is of significant importance from both a fundamental and practical perspective \cite{Nagaosa2010}. Conventionally, the AHE is considered proportional to magnetization and so large effects are not anticipated in antiferromagnets. However, over the last decade it has been appreciated that a spontaneous AHE can arise in antiferromagnets with zero net magnetisation that break certain symmetries, usually via the inclusion of spin-orbit coupling (SOC) \cite{Chen2014,Kubler2014,Zhang2017,Nakatsuji2015}. In addition to SOC a lack of inversion symmetry is also required, such as found in non-collinear antiferromagnetic (AFM) structures that arise from geometric frustration. In such scenarios, non-trivial electronic band structures lead to large values of the Berry curvature and, consequently, a large AHE. From a fundamental perspective, the AHE in antiferromagnets is of interest as the underlying physics also give rise to exotic quasiparticle excitations, such as the recently observed Dirac and Weyl fermions \cite{Ye2018,Kuroda2017}. From a technological perspective, the AHE offers novel opportunities in spintronics devices that utilize control of the AFM state despite zero magnetization \cite{Smejkal2018,Liu2018}.

Mn antiperovskites, with general formula Mn$_{3}A$N, have attracted a revival of interest due to a range of fascinating phenomena, such as temperature independent resistivity \cite{Chi2001}, abnormal thermal expansion  \cite{Takenaka2014}, and barocaloric effects \cite{Matsunami2014,Boldrin2018b}. These properties are underpinned by a non-collinear AFM structure and magnetovolume coupling that induces strong first-order character to the magnetic transitions. Moreover, the chemical flexibility of the structure allows many elements to be doped on the $A$ site, providing synthetic control of their functional properties \cite{Takenaka2014}. In thin film form the family are predicted to host giant piezomagnetism, a linear dependence of the magnetisation on the elastic stress tensor \cite{Zemen2017a,Zemen2017b}, a characteristic we have recently demonstrated experimentally in Mn$_{3}$NiN thin films \cite{Boldrin2018a}. Taken together, these materials offer a unique mixture of magnetic and structural properties and yet the magnetoconductance is a property that is as yet little examined \cite{Gurung2019}, particularly as a function of strain.

\begin{figure}
\includegraphics[width=0.45\textwidth]{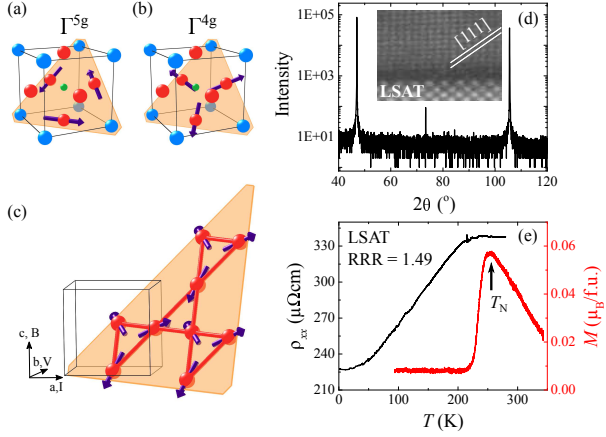}%
\caption{(a – b) The non-collinear AFM structures termed \Gfive\ and \Gfour, respectively. (c) The orientation of the kagome (111) plane relative to the cubic unit cell. The substrate is in the (110) plane. For magnetotransport, voltage and current axes are in the (110) plane, with the magnetic field applied along [001]. (d) Out-of-plane X-ray diffraction data and HAADF-STEM image (full scale 3.75nm) of a 50\,nm Mn$_{3}$NiN film grown on LSAT. (e) Longitudinal resistivity (left axis) and magnetization (right axis) measured out-of-plane as a function of temperature on the same film as in (d). The Ne\'{e}l temperature \TN\ and the residual resistance ratio (RRR) are indicated.}
\label{Fig1}
\end{figure}

We consider the $A =$ Ni member of the family, Mn$_{3}$NiN, studied previously in the bulk for its nearly-zero temperature dependent resistivity \cite{Chi2001}, above its strongly first-order paramagnetic to AFM transition at $T_{\mathrm{N}} = 260$\,K, where the material undergoes a 0.4\,\% volume change. In the AFM phase, the moments which reside on the Mn sites undergo a temperature dependent rotation between two high-symmetry structures on cooling, termed \Gfive\ and \Gfour\ (see Figs. \ref{Fig1}(a) and \ref{Fig1}(b), respectively) – both closely related to the magnetic structure found in the binary alloys Mn$_{3}A$ \cite{Yasukochi1961,Kren1971}. In this article we investigate the AHE in Mn$_{3}$NiN thin films grown on (LaAlO$_{3}$)$_{0.3}$(Sr$_{2}$TaAlO$_{6}$)$_{0.7}$ (LSAT) and SrTiO$_{3}$ (STO) substrates which produces a small tensile/compressive in-plane biaxial strain of the order of $\pm$0.1\,\%. We find the AHE neither scales directly with magnetization nor the electrical conductivity, as would be expected conventionally. Instead, we observe an enhanced AHE in the AFM state consistent with that expected for non-collinear antiferromagnets with an enhanced Berry phase contribution. The intrinsic AHE originated from the Berry phase can be attributed to the \Gfour\ magnetic configuration, consistent with the symmetry analysis \cite{Suzuki2017}, and we find that in-plane compressive strains enhance the AHE leading to a significant piezospintronic effect. 

Thin films of Mn$_{3}$NiN were grown as described in \cite{Boldrin2018a}. The sample primarily examined in this study is a 50\,nm thick Mn$_{3}$NiN film on a (001) LSAT substrate. The anomalous Hall conductivity is also compared to a second 100\,nm film on LSAT to study reproducibility, and to a 100\,nm film on (001) STO to examine the influence of strain. X-ray diffraction (XRD) data collected along the [00l] direction on the 50nm Mn$_{3}$NiN grown on LSAT is shown in Fig. \ref{Fig1}(d) along with high-angle annular dark field scanning transmission electron microscopy (HAADF-STEM). The XRD shows that the film peak cannot be distinguished from the substrate peak, and that the film therefore is strained relative to the bulk material by growth on the substrate. The HAADF-STEM shows a highly crystalline film, although with a slight rotation around the [110] direction relative to the substrate. Four terminal magnetotransport measurements were performed as a function of temperature using the standard van der Paauw method. The temperature dependent longitudinal resistivity, $\rho_{xx}$, is shown in Figs. 1(e) along with the temperature dependent magnetization performed using the VSM option of a Quantum Design Physical Property Measurement System (PPMS-9T). The residual resistance ratio of 1.5 is in line with structurally similar films \cite{Matsumoto2017}. The film shows a transition to an AFM state at $T_{\mathrm{N}} = 250$\,K.

In order to calculate the intrinsic anomalous Hall conductivity and the Berry phase contribution, the projector augmented wave method as implemented in the VASP code \cite{Kresse1993} was used, where the exchange correlation functional is formulated in the generalized gradient approximation as parameterized by Perdew-Burke-Ernzerhof \cite{Perdew1996}. Our results were obtained using a 13x13x13 $\Gamma$-centered k-mesh sampling and a 500\,eV energy cut-off to guarantee good convergence. The valence configurations of Mn, Ni and N, are $3d^{6}4s^{1}$, $3d^{8}4s^{2}$ and $2s^{2}2p^{3}$, respectively. The scheme in Ref. \cite{Mostofi2008} was followed so as to project the obtained DFT Bloch wave functions onto ‘maximally localized Wannier functions’ (MLWF). The MLWF \cite{Mostofi2008}, which were constructed for the $s$, $p$, and $d$ orbitals of the Mn and Ni atoms, and the $s$ and $p$ orbitals for N, generated on an $8 \times 8 \times 8$ k-mesh. The intrinsic anomalous Hall conductivity is evaluated by integrating the Berry curvature using the Wanniertools \cite{Wu2018} on a $500 \times 500 \times 500$ k-mesh, according to the formula \cite{Xiao2010}: 

\begin{equation}
\sigma_{\alpha,\beta} = \frac{-e^{2}}{\hbar} \int \frac{dk}{(2\pi^{3})} \sum\limits_{n(occ.)} f[\epsilon(k) - \mu]\Omega_{n,\alpha,\beta}(k)
\label{EQ1}
\end{equation}

\begin{equation}
\Omega_{n,\alpha,\beta}(k) = -2I \sum\limits_{m \neq n} \frac{\left\langle km | \nu_{\alpha}(k) | kn \right\rangle \left\langle kn | \nu_{\beta}(k) | km \right\rangle}{[\epsilon_{kn} - \epsilon_{km}]^{2}}
\label{EQ1}
\end{equation}

\noindent
where $f[\epsilon(k) - \mu]$ denotes the Fermi distribution function with Fermi energy indicated by $\mu$, and $\epsilon_{kn}$ ($\epsilon_{km}$) are the energy eigenvalues corresponding to occupied (unoccupied) Bloch band $n$ ($m$), $\nu_{\alpha}(k)$  ($\nu_{\beta}(k)$) corresponds to the velocity operator in Cartesian coordinates.

We turn now to the magnetotransport measurements performed at different temperatures above and below \TN. Regarding the measurement geometry, it is important to note that the kagome (111) plane that the moments reside in lies at $\sim$45 degrees to the film surface normal, as shown in Fig. \ref{Fig1}(c). In the Mn$_{3}A$ materials, the maximum AHE is observed when the applied field is in the kagome plane, and both voltage and current are orthogonal to this \cite{Nakatsuji2015}. In our thin films, we are unable to access this geometry. Instead, the applied field is applied normal to the film surface along $c$, with voltage and current directions both orthogonal to this axis and each other. 

\begin{figure}
\includegraphics[width=0.4\textwidth]{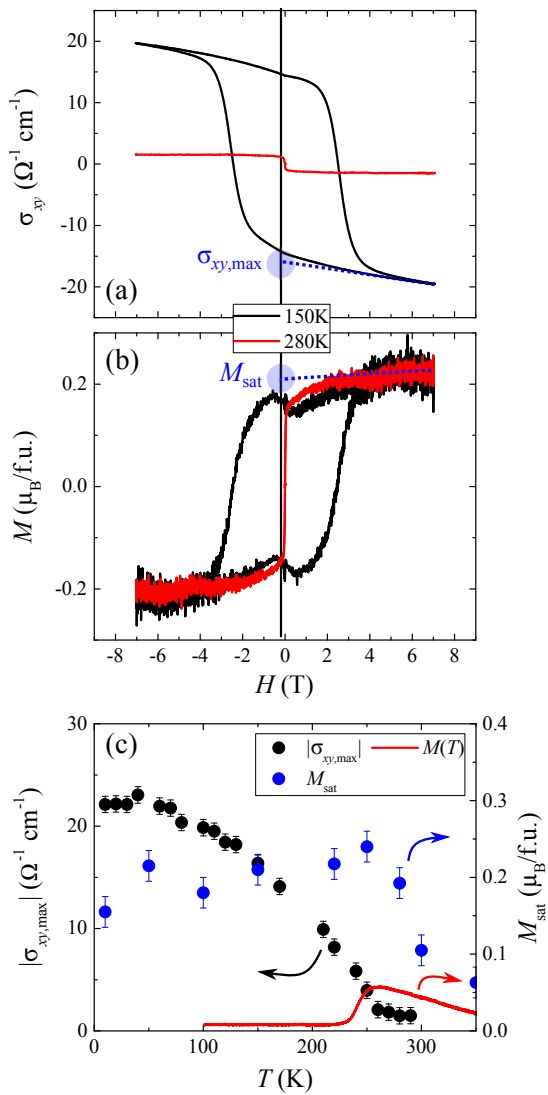}%
\caption{(a) Hall conductivity and out-of-plane magnetization measured as a function of field above and below \TN\ of the 50\,nm Mn$_{3}$NiN (LSAT) film. (b) Spontaneous Hall conductivity (left axis) and saturation magnetization, $M_{\mathrm{sat}}$, (right axis), extracted from the data in (a) and (b), respectively. The low field magnetization (ZFCW, $H = 0.05$\,T) is also shown for comparison.}
\label{Fig2}
\end{figure}

The Hall conductivity, $\sigma_{xy}$, is commonly defined as \sxy\ = -\pxy /\pxx$^{2}$ in the small Hall angle regime, \pxy\ $<<$ \pxx, which in our case \pxy /\pxx\ is smaller than 0.007 at all temperatures. Plots of \sxy$(H)$ at different temperatures for the 50\,nm film on LSAT are shown in Fig. \ref{Fig2}(a) and can be compared directly with magnetization measurements (out-of-plane) in the same field region in Fig. \ref{Fig2}(b). The shape of both Hall conductivity and magnetization are closely matched in the region shown above and below \TN. However, it is clear that the magnitude of \sxy\ is quite different above and below \TN\ whereas the magnitude of the magnetization remains roughly constant. 

Conventionally, measurements of transverse resistivity have contributions from the normal Hall effect proportional to the applied field, $R_{0}H$, and a term proportional to the magnetization, $R_{S}M$. As discussed, a further intrinsic term can contribute when a non-trivial electronic band structure causes a large Berry curvature. Deconvoluting these effects can be performed by observing their temperature dependencies \cite{Ye2018,Tian2009}. In Fig. \ref{Fig2}(c) we compare directly the temperature dependence of the Hall conductivity, $|$\sxy$|$, (where the normal Hall contribution has been removed, see Fig. \ref{Fig2}(a)), and the saturation magnetization $M_{\mathrm{sat}}$. For a simple collinear ferromagnet one would expect \sxy\ to scale with the magnetization. This relationship approximately holds between 300\,K and 250\,K where the magnetization increases linearly and the small but finite $|$\sxy$|$ behaves similarly. However, below \TN\ $|$\sxy$|$ increases steadily  and does not saturate until $T \sim 50$\,K, whereas the magnetization peaks at \TN\ and remains roughly constant upon further cooling. To check the reproducibility of this result, we performed further measurements on a 100\,nm thick film, again grown on LSAT. $|$\sxy$|$ shows a remarkably similar temperature dependence, although saturates at roughly half the value of the 50\,nm film at 10\,K (see Fig. \ref{FigS1}(a)). Evidently, the behaviour below \TN\ cannot be reconciled with the conventional description of the AHE being proportional to magnetization. 

\begin{figure}
\begin{center}
\includegraphics[width=0.45\textwidth]{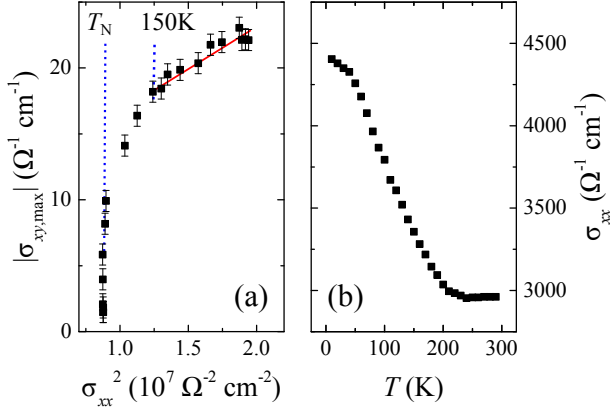}%
\end{center}
\caption{(a) A plot of the Hall conductivity, $|$\sxy$|$, as a function of the longitudinal conductivity, \sxx$^{2}$, and (b) the longitudinal resistivity, \sxx, as a function of temperature. The vertical blue lines in (a) indicate two distinct regions; (i) between \TN\ and $T = 150$\,K where \sxx\ remains relatively constant and (ii) below $T = 150$\,K where an upturn in \sxx\ is observed. The latter region has a roughly linear dependence between $|$\sxy$|$ and \sxx$^{2}$, indicated by the red line guiding the eye.}
\label{Fig3}
\end{figure}

Further understanding of the origins of the Hall conductivity can be gained through comparison with the longitudinal conductivity \cite{Ye2018,Tian2009}. Extrinsic contributions to the AHE are caused by scattering processes, which have similar temperature dependencies to the longitudinal conductivity. Thus, the scaling of both, in the absence of changes to magnetization, is given by the relationship \cite{Tian2009,Shitade2012}:

\begin{equation}
\sigma_{xy} = f(\sigma_{xx,0})\sigma_{xx}^{2} + \sigma_{xy}^{\mathrm{int}}
\label{EQ3}
\end{equation}

\noindent
where \sxxzero\ is the residual conductivity and \sxyint\ is the intrinsic anomalous Hall conductivity. A plot of $|$\sxy$|$ against \sxx$^{2}$ is shown in Fig. \ref{Fig3}(a) for the 50\,nm film on LSAT and the temperature dependence of \sxx\ is in Fig. \ref{Fig3}(b). Whilst the relatively small RRR of our film does not allow access to a large range of \sxx\ values as a function of temperature, certain trends can be seen in the data. Firstly, between \TN\ and $\sim$150\,K, where both the magnetization and longitudinal conductivity remain relatively constant (the latter changing by $<\,10$\,\%), $|$\sxy$|$ changes by an order of magnitude in a non-linear relationship with \sxx$^{2}$. Secondly, below 150\,K, \sxx\ begins to increase more rapidly, changing by $\sim$\,40\,\% by 10\,K, whilst $|$\sxy$|$ begins to plateau. In this region, $|$\sxy$|$ has a much more linear dependence with \sxx$^{2}$. This suggests the region has a contribution from extrinsic scattering-type processes, which is expected at low temperatures due to the increased carrier lifetime with larger \sxx. In the high temperature regime, the non-linear dependence of $|$\sxy$|$ with \sxx$^{2}$ is not consistent with extrinsic scattering processes. Thus, we conclude that this behaviour of $|$\sxy$|$ in this regime cannot be accounted for by either extrinsic scattering processes or changes in magnetization. 

Large intrinsic contributions to the AHE in Mn$_{3}A$ materials have recently been shown to arise from large Berry curvature \cite{Nakatsuji2015} due to the breaking of both time-reversal and certain spatial symmetries \cite{Chen2014}. We note that the magnetic structure of Mn$_{3}$NiN is remarkably similar to that found in Mn$_{3}$Ir and Mn$_{3}$Pt, where a large AHE has been predicted in the former and observed in the latter \cite{Chen2014,Liu2018}. This strongly suggests that the observed increase in AHE below \TN\ in Mn$_{3}$NiN is of similar origins to Mn$_{3}A$, \emph{i.e.} a large Berry curvature from the non-collinear antiferromagnetism. Two noteworthy differences are (i) in Mn$_{3}$NiN the moments have been rotated within the (111) plane between two different magnetic structures, \Gfour\ and \Gfive\ (See Fig. \ref{Fig1}(a-b)), and (ii) the thin films have a small moment induced by the piezomagnetic effect \cite{Boldrin2018a}. 

\begin{table}[!bh]
\begin{ruledtabular}
\begin{tabular}{ c | cc }
Magnetic Phase & Canted \Gfour\ & Canted \Gfive\ \\ \hline
Magnetic & \multirow{2}{*}{$C2^{\prime}/m^{\prime}$ (12.62)} & \multirow{2}{*}{$C2/m$ (12.58)} \\ 
Space Group & & \\ \hline
AHC tensor & $\left[ \begin{array}{ccc} 0 & \sigma_{xy} & -\sigma_{zx} \\ -\sigma_{xy} & 0 & \sigma_{zx} \\ \sigma_{zx} & -\sigma_{zx} & 0 \end{array} \right]$ & $\left[ \begin{array}{ccc} 0 & 0 & -\sigma_{zx} \\ 0 & 0 & -\sigma_{zx} \\ \sigma_{zx} & \sigma_{zx} & 0 \end{array} \right]$
\end{tabular}
\caption{\label{Table1} Symmetry imposed AHC tensor for the canted \Gfour\ and \Gfive\ states under strain.}
\end{ruledtabular}
\end{table}

\begin{figure}[!t]
\includegraphics[width=0.4\textwidth]{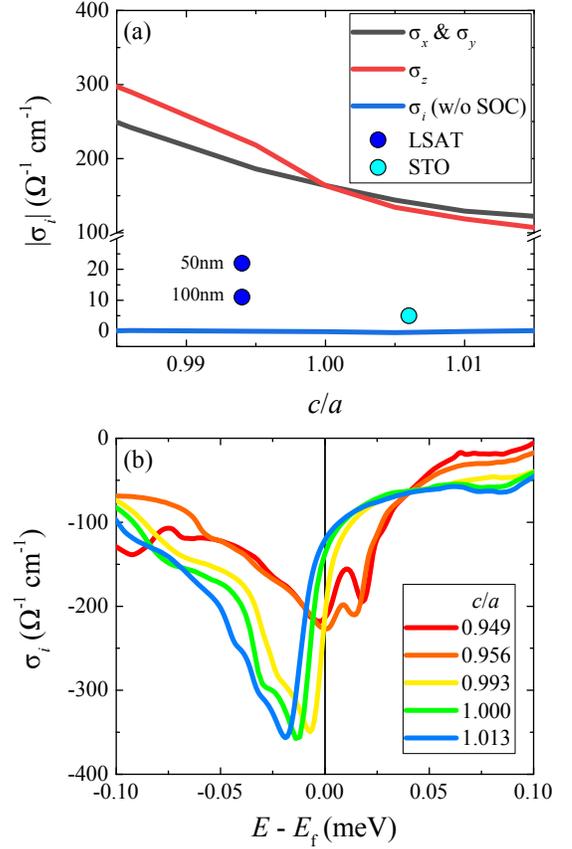}%
\caption{(a) The behaviour of the calculated intrinsic anomalous Hall conductivity components, $\sigma_{i}$, under strain (red, black and blue curves) in comparison to the experimental values (blue and cyan points). (b) The AHC for different strain values as a function of energy.}
\label{Fig4}
\end{figure}

The calculated anomalous Hall conductivity (AHC) in Mn$_{3}$NiN as a function of strain is shown in Fig. \ref{Fig4}(a), along with the experimentally measured values at 10\,K in the two LSAT films previously discussed. To compare the experimental strain dependence of the AHE with the predicted values we have also measured a 100\,nm Mn$_{3}$NiN film on an STO substrate (the preparation of which is described in \cite{Boldrin2018a}), also plotted in Fig \ref{Fig4}(a) and the full temperature dependence in Fig. \ref{FigS1}(b). It is clear that compressive (tensile) strain enhances (reduces) the intrinsic AHC, as obtained by DFT calculations assuming the \Gfour\ magnetic structure. It is noted that the unstrained \Gfive\ state is invariant under the application of $M_{\overline{1}10}$ and equivalent mirror planes. Such symmetry operations oddly transform the Berry curvature, which vanishes after integrating over the whole Brillouin zone \cite{Gurung2019}. Hence, all AHC components are zero. When finite strain is applied the symmetry is reduced to space group $C2/m$ (12.58), forcing only $\sigma_{z}$ to be zero (see Fig. \ref{FigS2}). However, the unstrained \Gfour\ state is invariant under the application of the combined $TM_{\overline{1}10}$ and equivalent symmetry operations. In this case, the Berry curvature is evenly transformed,
leading to finite AHC values \cite{Gurung2019,Suzuki2017}. In the strained case, the symmetry operations impose $\sigma_{x} = \sigma_{y} \neq 0$ and $\sigma_{z} \neq 0$. The form of the AHC tensor for both strained configurations was constructed using Ref. \cite{ZeleznyProgram,Smejkal} and it is presented in Table \ref{Table1}. Similar behavior of the AHC is predicted for unstrained Mn$_{3}$GaN \cite{Samathrakis2019,Gurung2019}. 
We note that the magnitude of the experimentally observed AHC is one order of magnitude smaller than the values from DFT calculations. This can be attributed to the possible mixture of the \Gfive\ and \Gfour\ states which are separated by an energy barrier, because the \Gfour\ state is less than 0.7\,meV lower in energy than the \Gfive\ state. Large variation in the AHC can also be caused by variation in  electronic doping (see Fig. \ref{Fig4}(b)), \emph{e.g.} a shift of the Fermi energy by 0.05\,eV leads to a change of 150\,S\,cm$^{-1}$ in the calculated conductivity value. Interestingly, as the anomalous Nernst effect is related to the derivative of AHC with respect to the Fermi energy, Fig. \ref{Fig4}(b) suggests it should be significant in these materials.

\begin{table*}[!th]
\caption{A comparison of the anomalous Hall effects present in a variety of magnetic materials. The final column parameterizes the strength of the AHE relative to the magnetization of the material.}
\begin{ruledtabular}
\begin{tabular}{ c | ccccc }
\multirow{2}{*}{Sample} & \multirow{2}{*}{$T$ (K)} & $M_\mathrm{sat}$  & $|$\sxy$|$ & \sxx\ & $|$\sxy$|$ per $M$ \\
& & ($\mu_{\mathrm{B}}$\,ion$^{-1}$) & ($\Omega^{-1}$\,cm$^{-1}$) & ($\Omega^{-1}$\,cm$^{-1}$) & ($\Omega^{-1}$\,cm$^{-1}$\,$\mu_{\mathrm{B}}^{-1}$\,ion) \\ \hline
Mn$_{3}$NiN (LSAT) & 10 & 0.053 & 22 & 4100 & 415 \\
Mn$_{3}$NiN (LSAT) & 300 & 0.033 & 1.5 & 3000 & 35 \\
Mn$_{3}$Sn \cite{Nakatsuji2015} & 100 & 0.003 & 100 & 3125 & 40000 \\
Mn$_{3}$Pt \cite{Liu2018} & 100 & 0.005 & 100 & 3300 & 20000 \\
Fe$_{3}$Sn$_{2}$ \cite{Ye2018} & 300 & 1.7 & 180 & 2000 & 105 \\
Mn$_{3}$CuN \cite{Matsumoto2017} & 5 & 0.23 & 4.2 & 2630 & 18 \\
Fe (2.5\,nm) \cite{Sangiao2009} & 300 & 2.2 & 300 & 20000 & 135 \\
Co$_{3}$Sn$_{2}$S$_{2}$ \cite{Wang2018} & 100 & 0.25 & 1000 & 7000 & 4000
\end{tabular}
\end{ruledtabular}
\label{Table2}
\end{table*}

The intrinsic contribution shown here relies on the finite spin-orbit coupling interaction in Mn$_{3}$NiN. The AHC is vanishingly small if the spin-orbit coupling is set to zero. This can be understood when considering that the magnetic moments of all atoms form a coplanar spin configuration and the $R_{S}T$ symmetry is preserved in the absence of spin-orbit coupling \cite{Suzuki2017}. This is indeed observed in our explicit calculations on both \Gfive\ and \Gfour\ configurations (see Fig \ref{Fig4}(a)). That is, we find that the topological Hall contribution in the co-planar spin geometry is insignificant. Furthermore, we suspect that there would be no net contribution to the AHC due to the Weyl points \cite{Kubler2014}. This is due to the fact that the Weyl points always appear in pairs with opposite chiralities and for magnetic materials they are supposed to be at the same energy \cite{Halasz2012}.

In Table \ref{Table2} we summarise the magnitude of the AHE in Mn$_{3}$NiN and compare to a number of other magnetic materials. Firstly, we see that \sxy\ is an order of magnitude larger in the AFM state despite only a doubling of the size of $M$. Compared to an Fe thin film, the relative strength of the AHE to $M$ in Mn$_{3}$NiN is around triple, further evidence that it is larger than expected. In antiperovskite Mn$_{3}$CuN, where the magnetic structure is a collinear antiferromagnet, the AHE is roughly 5 times smaller than for $A = $ Ni, and relative to its magnetization it is an order of magnitude smaller. In the purely AFM Mn$_{3}A$ materials \sxy\ is 5 times larger, whilst in the ferromagnetic Fe$_{3}$Sn$_{2}$ \cite{Kida2011,Ye2018} and Co$_{3}$Sn$_{2}$S$_{2}$ \cite{Wang2018} \sxy\ is 9 and 50 times larger, respectively; all these examples have large SOC from Sn or Pt which in turn increases the Berry curvature driven AHE. The Mn$_{3}A$N material family is chemically flexible, with the $A$ site taking many transition and post-transition metals, such that elements with large SOC may be incorporated. Therefore, our results should spur further experiments for an enlarged AHE with large SOC elements, but also theoretical work to reveal the underlying physics giving rise to these effects.

In conclusion, we have measured the AHE in strained thin films of Mn$_{3}$NiN and compared our results with theoretical prediction of the intrinsic effect for the \Gfour\ AFM phase in addition to its strain dependence. We find a large Berry curvature induced by the non-collinear AFM structure which is of similar origin as that found in the isostructural Mn$_{3}$Pt. Theoretical analysis demonstrates that the intrinsic AHE is highly sensitive to strain, doping and the ratio of \Gfive\ and \Gfour. Moreover, the AHE relies on spin-orbit coupling and therefore larger effects can be expected in the wider Mn$_{3}A$N family with heavier $A$ elements. The predicted strain dependence of the AHE, displays a sharp change in the AHC relative to the Fermi energy for compressively strained films indicating that a large anomalous Nernst effect is expected in these materials. Recent papers demonstrating manipulation of AFM structures by growth on piezoelectric substrates \cite{Liu2018} could inspire similar developments in Mn$_{3}A$N thin films where one could create large piezospintronic effects via strain coupling. 

%

\begin{acknowledgments}
We acknowledge support from the Henry Royce Institute made through EPSRC grant EP/R00661X/1 and EPSRC Impact Acceleration Account funding: Localised magnetic repository (LoMaRe) a high-performance nonvolatile memory device. We thank Jakub $\mathrm{\check{Z}}$elenzn$\mathrm{\acute{y}}$ for fruitful discussions of the form of the AHC tensor. The authors acknowledge support from the DFG-SPP 1666 programme and the Lichtenberg high performance computer of TU Darmstadt.
\end{acknowledgments}


\begin{thebibliography}{35}%
\makeatletter
\providecommand \@ifxundefined [1]{%
 \@ifx{#1\undefined}
}%
\providecommand \@ifnum [1]{%
 \ifnum #1\expandafter \@firstoftwo
 \else \expandafter \@secondoftwo
 \fi
}%
\providecommand \@ifx [1]{%
 \ifx #1\expandafter \@firstoftwo
 \else \expandafter \@secondoftwo
 \fi
}%
\providecommand \natexlab [1]{#1}%
\providecommand \enquote  [1]{``#1''}%
\providecommand \bibnamefont  [1]{#1}%
\providecommand \bibfnamefont [1]{#1}%
\providecommand \citenamefont [1]{#1}%
\providecommand \href@noop [0]{\@secondoftwo}%
\providecommand \href [0]{\begingroup \@sanitize@url \@href}%
\providecommand \@href[1]{\@@startlink{#1}\@@href}%
\providecommand \@@href[1]{\endgroup#1\@@endlink}%
\providecommand \@sanitize@url [0]{\catcode `\\12\catcode `\$12\catcode
  `\&12\catcode `\#12\catcode `\^12\catcode `\_12\catcode `\%12\relax}%
\providecommand \@@startlink[1]{}%
\providecommand \@@endlink[0]{}%
\providecommand \url  [0]{\begingroup\@sanitize@url \@url }%
\providecommand \@url [1]{\endgroup\@href {#1}{\urlprefix }}%
\providecommand \urlprefix  [0]{URL }%
\providecommand \Eprint [0]{\href }%
\providecommand \doibase [0]{http://dx.doi.org/}%
\providecommand \selectlanguage [0]{\@gobble}%
\providecommand \bibinfo  [0]{\@secondoftwo}%
\providecommand \bibfield  [0]{\@secondoftwo}%
\providecommand \translation [1]{[#1]}%
\providecommand \BibitemOpen [0]{}%
\providecommand \bibitemStop [0]{}%
\providecommand \bibitemNoStop [0]{.\EOS\space}%
\providecommand \EOS [0]{\spacefactor3000\relax}%
\providecommand \BibitemShut  [1]{\csname bibitem#1\endcsname}%
\let\auto@bib@innerbib\@empty
\bibitem [{\citenamefont {Nagaosa}\ \emph {et~al.}(2010)\citenamefont
  {Nagaosa}, \citenamefont {Sinova}, \citenamefont {Onoda}, \citenamefont
  {MacDonald},\ and\ \citenamefont {Ong}}]{Nagaosa2010}%
  \BibitemOpen
  \bibfield  {author} {\bibinfo {author} {\bibfnamefont {N.}~\bibnamefont
  {Nagaosa}}, \bibinfo {author} {\bibfnamefont {J.}~\bibnamefont {Sinova}},
  \bibinfo {author} {\bibfnamefont {S.}~\bibnamefont {Onoda}}, \bibinfo
  {author} {\bibfnamefont {A.~H.}\ \bibnamefont {MacDonald}}, \ and\ \bibinfo
  {author} {\bibfnamefont {N.~P.}\ \bibnamefont {Ong}},\ }\href@noop {}
  {\bibfield  {journal} {\bibinfo  {journal} {Reviews of Modern Physics}\
  }\textbf {\bibinfo {volume} {82}},\ \bibinfo {pages} {1539} (\bibinfo {year}
  {2010})}\BibitemShut {NoStop}%
\bibitem [{\citenamefont {Chen}\ \emph {et~al.}(2014)\citenamefont {Chen},
  \citenamefont {Niu},\ and\ \citenamefont {Macdonald}}]{Chen2014}%
  \BibitemOpen
  \bibfield  {author} {\bibinfo {author} {\bibfnamefont {H.}~\bibnamefont
  {Chen}}, \bibinfo {author} {\bibfnamefont {Q.}~\bibnamefont {Niu}}, \ and\
  \bibinfo {author} {\bibfnamefont {A.~H.}\ \bibnamefont {Macdonald}},\ }\href
  {\doibase 10.1103/PhysRevLett.112.017205} {\bibfield  {journal} {\bibinfo
  {journal} {Physical Review Letters}\ }\textbf {\bibinfo {volume} {112}},\
  \bibinfo {pages} {017205} (\bibinfo {year} {2014})}\BibitemShut {NoStop}%
\bibitem [{\citenamefont {K{\"{u}}bler}\ and\ \citenamefont
  {Felser}(2014)}]{Kubler2014}%
  \BibitemOpen
  \bibfield  {author} {\bibinfo {author} {\bibfnamefont {J.}~\bibnamefont
  {K{\"{u}}bler}}\ and\ \bibinfo {author} {\bibfnamefont {C.}~\bibnamefont
  {Felser}},\ }\href {\doibase 10.1209/0295-5075/108/67001} {\bibfield
  {journal} {\bibinfo  {journal} {Europhysics Letters}\ }\textbf {\bibinfo
  {volume} {108}},\ \bibinfo {pages} {67001} (\bibinfo {year}
  {2014})}\BibitemShut {NoStop}%
\bibitem [{\citenamefont {Zhang}\ \emph {et~al.}(2017)\citenamefont {Zhang},
  \citenamefont {Sun}, \citenamefont {Yang}, \citenamefont
  {{\v{Z}}elezn{\'{y}}}, \citenamefont {Parkin}, \citenamefont {Felser},\ and\
  \citenamefont {Yan}}]{Zhang2017}%
  \BibitemOpen
  \bibfield  {author} {\bibinfo {author} {\bibfnamefont {Y.}~\bibnamefont
  {Zhang}}, \bibinfo {author} {\bibfnamefont {Y.}~\bibnamefont {Sun}}, \bibinfo
  {author} {\bibfnamefont {H.}~\bibnamefont {Yang}}, \bibinfo {author}
  {\bibfnamefont {J.}~\bibnamefont {{\v{Z}}elezn{\'{y}}}}, \bibinfo {author}
  {\bibfnamefont {S.~P.}\ \bibnamefont {Parkin}}, \bibinfo {author}
  {\bibfnamefont {C.}~\bibnamefont {Felser}}, \ and\ \bibinfo {author}
  {\bibfnamefont {B.}~\bibnamefont {Yan}},\ }\href {\doibase
  10.1103/PhysRevB.95.075128} {\bibfield  {journal} {\bibinfo  {journal}
  {Physical Review B}\ }\textbf {\bibinfo {volume} {95}},\ \bibinfo {pages}
  {075128} (\bibinfo {year} {2017})}\BibitemShut {NoStop}%
\bibitem [{\citenamefont {Nakatsuji}\ \emph {et~al.}(2015)\citenamefont
  {Nakatsuji}, \citenamefont {Kiyohara},\ and\ \citenamefont
  {Higo}}]{Nakatsuji2015}%
  \BibitemOpen
  \bibfield  {author} {\bibinfo {author} {\bibfnamefont {S.}~\bibnamefont
  {Nakatsuji}}, \bibinfo {author} {\bibfnamefont {N.}~\bibnamefont {Kiyohara}},
  \ and\ \bibinfo {author} {\bibfnamefont {T.}~\bibnamefont {Higo}},\ }\href
  {\doibase 10.1038/nature15723} {\bibfield  {journal} {\bibinfo  {journal}
  {Nature}\ }\textbf {\bibinfo {volume} {527}},\ \bibinfo {pages} {212}
  (\bibinfo {year} {2015})}\BibitemShut {NoStop}%
\bibitem [{\citenamefont {Ye}\ \emph {et~al.}(2018)\citenamefont {Ye},
  \citenamefont {Kang}, \citenamefont {Liu}, \citenamefont {Cube},
  \citenamefont {Wicker}, \citenamefont {Suzuki}, \citenamefont {Jozwiak},
  \citenamefont {Bostwick}, \citenamefont {Rotenberg}, \citenamefont {Bell},
  \citenamefont {Fu}, \citenamefont {Comin},\ and\ \citenamefont
  {Checkelsky}}]{Ye2018}%
  \BibitemOpen
  \bibfield  {author} {\bibinfo {author} {\bibfnamefont {L.}~\bibnamefont
  {Ye}}, \bibinfo {author} {\bibfnamefont {M.}~\bibnamefont {Kang}}, \bibinfo
  {author} {\bibfnamefont {J.}~\bibnamefont {Liu}}, \bibinfo {author}
  {\bibfnamefont {F.~V.}\ \bibnamefont {Cube}}, \bibinfo {author}
  {\bibfnamefont {C.~R.}\ \bibnamefont {Wicker}}, \bibinfo {author}
  {\bibfnamefont {T.}~\bibnamefont {Suzuki}}, \bibinfo {author} {\bibfnamefont
  {C.}~\bibnamefont {Jozwiak}}, \bibinfo {author} {\bibfnamefont
  {A.}~\bibnamefont {Bostwick}}, \bibinfo {author} {\bibfnamefont
  {E.}~\bibnamefont {Rotenberg}}, \bibinfo {author} {\bibfnamefont {D.~C.}\
  \bibnamefont {Bell}}, \bibinfo {author} {\bibfnamefont {L.}~\bibnamefont
  {Fu}}, \bibinfo {author} {\bibfnamefont {R.}~\bibnamefont {Comin}}, \ and\
  \bibinfo {author} {\bibfnamefont {J.~G.}\ \bibnamefont {Checkelsky}},\ }\href
  {\doibase 10.1038/nature25987} {\bibfield  {journal} {\bibinfo  {journal}
  {Nature}\ }\textbf {\bibinfo {volume} {555}},\ \bibinfo {pages} {638}
  (\bibinfo {year} {2018})}\BibitemShut {NoStop}%
\bibitem [{\citenamefont {Kuroda}\ \emph {et~al.}(2017)\citenamefont {Kuroda},
  \citenamefont {Tomita}, \citenamefont {Suzuki}, \citenamefont {Bareille},
  \citenamefont {Nugroho}, \citenamefont {Goswami}, \citenamefont {Ochi},
  \citenamefont {Ikhlas}, \citenamefont {Nakayama}, \citenamefont {Akebi},
  \citenamefont {Noguchi}, \citenamefont {Ishii}, \citenamefont {Inami},
  \citenamefont {Ono}, \citenamefont {Kumigashira}, \citenamefont {Varykhalov},
  \citenamefont {Muro}, \citenamefont {Koretsune}, \citenamefont {Arita},
  \citenamefont {Shin}, \citenamefont {Kondo},\ and\ \citenamefont
  {Nakatsuji}}]{Kuroda2017}%
  \BibitemOpen
  \bibfield  {author} {\bibinfo {author} {\bibfnamefont {K.}~\bibnamefont
  {Kuroda}}, \bibinfo {author} {\bibfnamefont {T.}~\bibnamefont {Tomita}},
  \bibinfo {author} {\bibfnamefont {M.~T.}\ \bibnamefont {Suzuki}}, \bibinfo
  {author} {\bibfnamefont {C.}~\bibnamefont {Bareille}}, \bibinfo {author}
  {\bibfnamefont {A.~A.}\ \bibnamefont {Nugroho}}, \bibinfo {author}
  {\bibfnamefont {P.}~\bibnamefont {Goswami}}, \bibinfo {author} {\bibfnamefont
  {M.}~\bibnamefont {Ochi}}, \bibinfo {author} {\bibfnamefont {M.}~\bibnamefont
  {Ikhlas}}, \bibinfo {author} {\bibfnamefont {M.}~\bibnamefont {Nakayama}},
  \bibinfo {author} {\bibfnamefont {S.}~\bibnamefont {Akebi}}, \bibinfo
  {author} {\bibfnamefont {R.}~\bibnamefont {Noguchi}}, \bibinfo {author}
  {\bibfnamefont {R.}~\bibnamefont {Ishii}}, \bibinfo {author} {\bibfnamefont
  {N.}~\bibnamefont {Inami}}, \bibinfo {author} {\bibfnamefont
  {K.}~\bibnamefont {Ono}}, \bibinfo {author} {\bibfnamefont {H.}~\bibnamefont
  {Kumigashira}}, \bibinfo {author} {\bibfnamefont {A.}~\bibnamefont
  {Varykhalov}}, \bibinfo {author} {\bibfnamefont {T.}~\bibnamefont {Muro}},
  \bibinfo {author} {\bibfnamefont {T.}~\bibnamefont {Koretsune}}, \bibinfo
  {author} {\bibfnamefont {R.}~\bibnamefont {Arita}}, \bibinfo {author}
  {\bibfnamefont {S.}~\bibnamefont {Shin}}, \bibinfo {author} {\bibfnamefont
  {T.}~\bibnamefont {Kondo}}, \ and\ \bibinfo {author} {\bibfnamefont
  {S.}~\bibnamefont {Nakatsuji}},\ }\href {\doibase 10.1038/nmat4987}
  {\bibfield  {journal} {\bibinfo  {journal} {Nature Materials}\ }\textbf
  {\bibinfo {volume} {16}},\ \bibinfo {pages} {1090} (\bibinfo {year}
  {2017})}\BibitemShut {NoStop}%
\bibitem [{\citenamefont {{\v{S}}mejkal}\ \emph {et~al.}(2018)\citenamefont
  {{\v{S}}mejkal}, \citenamefont {Mokrousov}, \citenamefont {Yan},\ and\
  \citenamefont {MacDonald}}]{Smejkal2018}%
  \BibitemOpen
  \bibfield  {author} {\bibinfo {author} {\bibfnamefont {L.}~\bibnamefont
  {{\v{S}}mejkal}}, \bibinfo {author} {\bibfnamefont {Y.}~\bibnamefont
  {Mokrousov}}, \bibinfo {author} {\bibfnamefont {B.}~\bibnamefont {Yan}}, \
  and\ \bibinfo {author} {\bibfnamefont {A.~H.}\ \bibnamefont {MacDonald}},\
  }\href {\doibase 10.1038/s41567-018-0064-5} {\bibfield  {journal} {\bibinfo
  {journal} {Nature Physics}\ }\textbf {\bibinfo {volume} {14}},\ \bibinfo
  {pages} {242} (\bibinfo {year} {2018})}\BibitemShut {NoStop}%
\bibitem [{\citenamefont {Liu}\ \emph {et~al.}(2018)\citenamefont {Liu},
  \citenamefont {Chen}, \citenamefont {Wang}, \citenamefont {Liu},
  \citenamefont {Wang}, \citenamefont {Feng}, \citenamefont {Yan},
  \citenamefont {Wang}, \citenamefont {Jiang}, \citenamefont {Coey},\ and\
  \citenamefont {MacDonald}}]{Liu2018}%
  \BibitemOpen
  \bibfield  {author} {\bibinfo {author} {\bibfnamefont {Z.~Q.}\ \bibnamefont
  {Liu}}, \bibinfo {author} {\bibfnamefont {H.}~\bibnamefont {Chen}}, \bibinfo
  {author} {\bibfnamefont {J.~M.}\ \bibnamefont {Wang}}, \bibinfo {author}
  {\bibfnamefont {J.~H.}\ \bibnamefont {Liu}}, \bibinfo {author} {\bibfnamefont
  {K.}~\bibnamefont {Wang}}, \bibinfo {author} {\bibfnamefont {Z.~X.}\
  \bibnamefont {Feng}}, \bibinfo {author} {\bibfnamefont {H.}~\bibnamefont
  {Yan}}, \bibinfo {author} {\bibfnamefont {X.~R.}\ \bibnamefont {Wang}},
  \bibinfo {author} {\bibfnamefont {C.~B.}\ \bibnamefont {Jiang}}, \bibinfo
  {author} {\bibfnamefont {J.~M.~D.}\ \bibnamefont {Coey}}, \ and\ \bibinfo
  {author} {\bibfnamefont {A.~H.}\ \bibnamefont {MacDonald}},\ }\href {\doibase
  10.1038/s41928-018-0040-1} {\bibfield  {journal} {\bibinfo  {journal} {Nature
  Electronics}\ }\textbf {\bibinfo {volume} {1}},\ \bibinfo {pages} {172}
  (\bibinfo {year} {2018})}\BibitemShut {NoStop}%
\bibitem [{\citenamefont {Chi}\ \emph {et~al.}(2001)\citenamefont {Chi},
  \citenamefont {Kim},\ and\ \citenamefont {Hur}}]{Chi2001}%
  \BibitemOpen
  \bibfield  {author} {\bibinfo {author} {\bibfnamefont {E.~O.}\ \bibnamefont
  {Chi}}, \bibinfo {author} {\bibfnamefont {W.~S.}\ \bibnamefont {Kim}}, \ and\
  \bibinfo {author} {\bibfnamefont {N.~H.}\ \bibnamefont {Hur}},\ }\href
  {\doibase 10.1016/S0038-1098(01)00395-7} {\bibfield  {journal} {\bibinfo
  {journal} {Solid State Communications}\ }\textbf {\bibinfo {volume} {120}},\
  \bibinfo {pages} {307} (\bibinfo {year} {2001})}\BibitemShut {NoStop}%
\bibitem [{\citenamefont {Takenaka}\ \emph {et~al.}(2014)\citenamefont
  {Takenaka}, \citenamefont {Ichigo}, \citenamefont {Hamada}, \citenamefont
  {Ozawa}, \citenamefont {Shibayama}, \citenamefont {Inagaki},\ and\
  \citenamefont {Asano}}]{Takenaka2014}%
  \BibitemOpen
  \bibfield  {author} {\bibinfo {author} {\bibfnamefont {K.}~\bibnamefont
  {Takenaka}}, \bibinfo {author} {\bibfnamefont {M.}~\bibnamefont {Ichigo}},
  \bibinfo {author} {\bibfnamefont {T.}~\bibnamefont {Hamada}}, \bibinfo
  {author} {\bibfnamefont {A.}~\bibnamefont {Ozawa}}, \bibinfo {author}
  {\bibfnamefont {T.}~\bibnamefont {Shibayama}}, \bibinfo {author}
  {\bibfnamefont {T.}~\bibnamefont {Inagaki}}, \ and\ \bibinfo {author}
  {\bibfnamefont {K.}~\bibnamefont {Asano}},\ }\href {\doibase
  10.1088/1468-6996/15/1/015009} {\bibfield  {journal} {\bibinfo  {journal}
  {Science and Technology of Advanced Materials}\ }\textbf {\bibinfo {volume}
  {15}},\ \bibinfo {pages} {15009} (\bibinfo {year} {2014})}\BibitemShut
  {NoStop}%
\bibitem [{\citenamefont {Matsunami}\ \emph {et~al.}(2014)\citenamefont
  {Matsunami}, \citenamefont {Fujita}, \citenamefont {Takenaka},\ and\
  \citenamefont {Kano}}]{Matsunami2014}%
  \BibitemOpen
  \bibfield  {author} {\bibinfo {author} {\bibfnamefont {D.}~\bibnamefont
  {Matsunami}}, \bibinfo {author} {\bibfnamefont {A.}~\bibnamefont {Fujita}},
  \bibinfo {author} {\bibfnamefont {K.}~\bibnamefont {Takenaka}}, \ and\
  \bibinfo {author} {\bibfnamefont {M.}~\bibnamefont {Kano}},\ }\href {\doibase
  10.1038/NMAT4117} {\bibfield  {journal} {\bibinfo  {journal} {Nature
  Materials}\ }\textbf {\bibinfo {volume} {14}},\ \bibinfo {pages} {73}
  (\bibinfo {year} {2014})}\BibitemShut {NoStop}%
\bibitem [{\citenamefont {Boldrin}\ \emph
  {et~al.}(2018{\natexlab{a}})\citenamefont {Boldrin}, \citenamefont
  {Mendive-Tapia}, \citenamefont {Zemen}, \citenamefont {Staunton},
  \citenamefont {Hansen}, \citenamefont {Aznar}, \citenamefont {Tamarit},
  \citenamefont {Barrio}, \citenamefont {Lloveras}, \citenamefont {Kim},
  \citenamefont {Moya},\ and\ \citenamefont {Cohen}}]{Boldrin2018b}%
  \BibitemOpen
  \bibfield  {author} {\bibinfo {author} {\bibfnamefont {D.}~\bibnamefont
  {Boldrin}}, \bibinfo {author} {\bibfnamefont {E.}~\bibnamefont
  {Mendive-Tapia}}, \bibinfo {author} {\bibfnamefont {J.}~\bibnamefont
  {Zemen}}, \bibinfo {author} {\bibfnamefont {J.~B.}\ \bibnamefont {Staunton}},
  \bibinfo {author} {\bibfnamefont {T.}~\bibnamefont {Hansen}}, \bibinfo
  {author} {\bibfnamefont {A.}~\bibnamefont {Aznar}}, \bibinfo {author}
  {\bibfnamefont {J.-L.}\ \bibnamefont {Tamarit}}, \bibinfo {author}
  {\bibfnamefont {M.}~\bibnamefont {Barrio}}, \bibinfo {author} {\bibfnamefont
  {P.}~\bibnamefont {Lloveras}}, \bibinfo {author} {\bibfnamefont
  {J.}~\bibnamefont {Kim}}, \bibinfo {author} {\bibfnamefont {X.}~\bibnamefont
  {Moya}}, \ and\ \bibinfo {author} {\bibfnamefont {L.~F.}\ \bibnamefont
  {Cohen}},\ }\href {\doibase 10.1103/PhysRevX.8.041035} {\bibfield  {journal}
  {\bibinfo  {journal} {Physical Review X}\ }\textbf {\bibinfo {volume} {8}},\
  \bibinfo {pages} {041035} (\bibinfo {year} {2018}{\natexlab{a}})}\BibitemShut
  {NoStop}%
\bibitem [{\citenamefont {Zemen}\ \emph
  {et~al.}(2017{\natexlab{a}})\citenamefont {Zemen}, \citenamefont {Gercsi},\
  and\ \citenamefont {Sandeman}}]{Zemen2017a}%
  \BibitemOpen
  \bibfield  {author} {\bibinfo {author} {\bibfnamefont {J.}~\bibnamefont
  {Zemen}}, \bibinfo {author} {\bibfnamefont {Z.}~\bibnamefont {Gercsi}}, \
  and\ \bibinfo {author} {\bibfnamefont {K.~G.}\ \bibnamefont {Sandeman}},\
  }\href {\doibase 10.1103/PhysRevB.96.024451} {\bibfield  {journal} {\bibinfo
  {journal} {Physical Review B}\ }\textbf {\bibinfo {volume} {96}},\ \bibinfo
  {pages} {024451} (\bibinfo {year} {2017}{\natexlab{a}})}\BibitemShut
  {NoStop}%
\bibitem [{\citenamefont {Zemen}\ \emph
  {et~al.}(2017{\natexlab{b}})\citenamefont {Zemen}, \citenamefont
  {Mendive-Tapia}, \citenamefont {Gercsi}, \citenamefont {Banerjee},
  \citenamefont {Staunton},\ and\ \citenamefont {Sandeman}}]{Zemen2017b}%
  \BibitemOpen
  \bibfield  {author} {\bibinfo {author} {\bibfnamefont {J.}~\bibnamefont
  {Zemen}}, \bibinfo {author} {\bibfnamefont {E.}~\bibnamefont
  {Mendive-Tapia}}, \bibinfo {author} {\bibfnamefont {Z.}~\bibnamefont
  {Gercsi}}, \bibinfo {author} {\bibfnamefont {R.}~\bibnamefont {Banerjee}},
  \bibinfo {author} {\bibfnamefont {J.~B.}\ \bibnamefont {Staunton}}, \ and\
  \bibinfo {author} {\bibfnamefont {K.~G.}\ \bibnamefont {Sandeman}},\ }\href
  {\doibase 10.1103/PhysRevB.95.184438} {\bibfield  {journal} {\bibinfo
  {journal} {Physical Review B}\ }\textbf {\bibinfo {volume} {95}},\ \bibinfo
  {pages} {184438} (\bibinfo {year} {2017}{\natexlab{b}})}\BibitemShut
  {NoStop}%
\bibitem [{\citenamefont {Boldrin}\ \emph
  {et~al.}(2018{\natexlab{b}})\citenamefont {Boldrin}, \citenamefont {Mihai},
  \citenamefont {Zou}, \citenamefont {Zemen}, \citenamefont {Thompson},
  \citenamefont {Ware}, \citenamefont {Neamtu}, \citenamefont {Ghivelder},
  \citenamefont {Esser}, \citenamefont {McComb}, \citenamefont {Petrov},\ and\
  \citenamefont {Cohen}}]{Boldrin2018a}%
  \BibitemOpen
  \bibfield  {author} {\bibinfo {author} {\bibfnamefont {D.}~\bibnamefont
  {Boldrin}}, \bibinfo {author} {\bibfnamefont {A.~P.}\ \bibnamefont {Mihai}},
  \bibinfo {author} {\bibfnamefont {B.}~\bibnamefont {Zou}}, \bibinfo {author}
  {\bibfnamefont {J.}~\bibnamefont {Zemen}}, \bibinfo {author} {\bibfnamefont
  {R.}~\bibnamefont {Thompson}}, \bibinfo {author} {\bibfnamefont
  {E.}~\bibnamefont {Ware}}, \bibinfo {author} {\bibfnamefont {B.~V.}\
  \bibnamefont {Neamtu}}, \bibinfo {author} {\bibfnamefont {L.}~\bibnamefont
  {Ghivelder}}, \bibinfo {author} {\bibfnamefont {B.}~\bibnamefont {Esser}},
  \bibinfo {author} {\bibfnamefont {D.~W.}\ \bibnamefont {McComb}}, \bibinfo
  {author} {\bibfnamefont {P.}~\bibnamefont {Petrov}}, \ and\ \bibinfo {author}
  {\bibfnamefont {L.~F.}\ \bibnamefont {Cohen}},\ }\href {\doibase
  10.1021/acsami.8b03112} {\bibfield  {journal} {\bibinfo  {journal} {ACS
  Applied Materials and Interfaces}\ }\textbf {\bibinfo {volume} {10}},\
  \bibinfo {pages} {18863} (\bibinfo {year} {2018}{\natexlab{b}})}\BibitemShut
  {NoStop}%
\bibitem [{\citenamefont {Gurung}\ \emph {et~al.}(2019)\citenamefont {Gurung},
  \citenamefont {Shao}, \citenamefont {Paudel},\ and\ \citenamefont
  {Tsymbal}}]{Gurung2019}%
  \BibitemOpen
  \bibfield  {author} {\bibinfo {author} {\bibfnamefont {G.}~\bibnamefont
  {Gurung}}, \bibinfo {author} {\bibfnamefont {D.-F.}\ \bibnamefont {Shao}},
  \bibinfo {author} {\bibfnamefont {T.~R.}\ \bibnamefont {Paudel}}, \ and\
  \bibinfo {author} {\bibfnamefont {E.~Y.}\ \bibnamefont {Tsymbal}},\ }\href
  {\doibase arXiv:1901.05040v1} {\  (\bibinfo {year} {2019}),\
  arXiv:1901.05040v1}\BibitemShut {NoStop}%
\bibitem [{\citenamefont {Yasukochi}\ \emph {et~al.}(1961)\citenamefont
  {Yasukochi}, \citenamefont {Kanematsu},\ and\ \citenamefont
  {Ohoyama}}]{Yasukochi1961}%
  \BibitemOpen
  \bibfield  {author} {\bibinfo {author} {\bibfnamefont {K.}~\bibnamefont
  {Yasukochi}}, \bibinfo {author} {\bibfnamefont {K.}~\bibnamefont
  {Kanematsu}}, \ and\ \bibinfo {author} {\bibfnamefont {T.}~\bibnamefont
  {Ohoyama}},\ }\href {\doibase 10.1143/JPSJ.16.1123} {\bibfield  {journal}
  {\bibinfo  {journal} {Journal of the Physical Society of Japan}\ }\textbf
  {\bibinfo {volume} {16}},\ \bibinfo {pages} {1123} (\bibinfo {year}
  {1961})}\BibitemShut {NoStop}%
\bibitem [{\citenamefont {Kr{\'{e}}n}\ \emph {et~al.}(1971)\citenamefont
  {Kr{\'{e}}n}, \citenamefont {K{\'{a}}d{\'{a}}r}, \citenamefont {P{\'{a}}l},
  \citenamefont {Zsoldos}, \citenamefont {Barberon},\ and\ \citenamefont
  {Fruchart}}]{Kren1971}%
  \BibitemOpen
  \bibfield  {author} {\bibinfo {author} {\bibfnamefont {E.}~\bibnamefont
  {Kr{\'{e}}n}}, \bibinfo {author} {\bibfnamefont {G.}~\bibnamefont
  {K{\'{a}}d{\'{a}}r}}, \bibinfo {author} {\bibfnamefont {L.}~\bibnamefont
  {P{\'{a}}l}}, \bibinfo {author} {\bibfnamefont {E.}~\bibnamefont {Zsoldos}},
  \bibinfo {author} {\bibfnamefont {M.}~\bibnamefont {Barberon}}, \ and\
  \bibinfo {author} {\bibfnamefont {R.}~\bibnamefont {Fruchart}},\ }\href
  {\doibase 10.1051/jphyscol:19711348} {\bibfield  {journal} {\bibinfo
  {journal} {Le Journal de Physique Colloques}\ }\textbf {\bibinfo {volume}
  {32}},\ \bibinfo {pages} {C1} (\bibinfo {year} {1971})}\BibitemShut {NoStop}%
\bibitem [{\citenamefont {Suzuki}\ \emph {et~al.}(2017)\citenamefont {Suzuki},
  \citenamefont {Koretsune}, \citenamefont {Ochi},\ and\ \citenamefont
  {Arita}}]{Suzuki2017}%
  \BibitemOpen
  \bibfield  {author} {\bibinfo {author} {\bibfnamefont {M.~T.}\ \bibnamefont
  {Suzuki}}, \bibinfo {author} {\bibfnamefont {T.}~\bibnamefont {Koretsune}},
  \bibinfo {author} {\bibfnamefont {M.}~\bibnamefont {Ochi}}, \ and\ \bibinfo
  {author} {\bibfnamefont {R.}~\bibnamefont {Arita}},\ }\href {\doibase
  10.1103/PhysRevB.95.094406} {\bibfield  {journal} {\bibinfo  {journal}
  {Physical Review B}\ }\textbf {\bibinfo {volume} {95}},\ \bibinfo {pages}
  {094406} (\bibinfo {year} {2017})}\BibitemShut {NoStop}%
\bibitem [{\citenamefont {Matsumoto}\ \emph {et~al.}(2017)\citenamefont
  {Matsumoto}, \citenamefont {Hatano}, \citenamefont {Urata}, \citenamefont
  {Iida}, \citenamefont {Takenaka},\ and\ \citenamefont
  {Ikuta}}]{Matsumoto2017}%
  \BibitemOpen
  \bibfield  {author} {\bibinfo {author} {\bibfnamefont {T.}~\bibnamefont
  {Matsumoto}}, \bibinfo {author} {\bibfnamefont {T.}~\bibnamefont {Hatano}},
  \bibinfo {author} {\bibfnamefont {T.}~\bibnamefont {Urata}}, \bibinfo
  {author} {\bibfnamefont {K.}~\bibnamefont {Iida}}, \bibinfo {author}
  {\bibfnamefont {K.}~\bibnamefont {Takenaka}}, \ and\ \bibinfo {author}
  {\bibfnamefont {H.}~\bibnamefont {Ikuta}},\ }\href {\doibase
  10.1103/PhysRevB.96.205153} {\bibfield  {journal} {\bibinfo  {journal}
  {Physical Review B}\ }\textbf {\bibinfo {volume} {96}},\ \bibinfo {pages}
  {205153} (\bibinfo {year} {2017})}\BibitemShut {NoStop}%
\bibitem [{\citenamefont {Kresse}\ and\ \citenamefont
  {Hafner}(1993)}]{Kresse1993}%
  \BibitemOpen
  \bibfield  {author} {\bibinfo {author} {\bibfnamefont {G.}~\bibnamefont
  {Kresse}}\ and\ \bibinfo {author} {\bibfnamefont {J.}~\bibnamefont
  {Hafner}},\ }\href {\doibase 10.1103/PhysRevB.47.558} {\bibfield  {journal}
  {\bibinfo  {journal} {Physical Review B}\ }\textbf {\bibinfo {volume} {47}},\
  \bibinfo {pages} {558} (\bibinfo {year} {1993})}\BibitemShut {NoStop}%
\bibitem [{\citenamefont {Perdew}\ \emph {et~al.}(1996)\citenamefont {Perdew},
  \citenamefont {Burke},\ and\ \citenamefont {Ernzerhof}}]{Perdew1996}%
  \BibitemOpen
  \bibfield  {author} {\bibinfo {author} {\bibfnamefont {J.~P.}\ \bibnamefont
  {Perdew}}, \bibinfo {author} {\bibfnamefont {K.}~\bibnamefont {Burke}}, \
  and\ \bibinfo {author} {\bibfnamefont {M.}~\bibnamefont {Ernzerhof}},\ }\href
  {\doibase 10.1103/PhysRevLett.77.3865} {\bibfield  {journal} {\bibinfo
  {journal} {Physical Review Letters}\ }\textbf {\bibinfo {volume} {77}},\
  \bibinfo {pages} {3865} (\bibinfo {year} {1996})}\BibitemShut {NoStop}%
\bibitem [{\citenamefont {Mostofi}\ \emph {et~al.}(2008)\citenamefont
  {Mostofi}, \citenamefont {Yates}, \citenamefont {Lee}, \citenamefont {Souza},
  \citenamefont {Vanderbilt},\ and\ \citenamefont {Marzari}}]{Mostofi2008}%
  \BibitemOpen
  \bibfield  {author} {\bibinfo {author} {\bibfnamefont {A.~A.}\ \bibnamefont
  {Mostofi}}, \bibinfo {author} {\bibfnamefont {J.~R.}\ \bibnamefont {Yates}},
  \bibinfo {author} {\bibfnamefont {Y.~S.}\ \bibnamefont {Lee}}, \bibinfo
  {author} {\bibfnamefont {I.}~\bibnamefont {Souza}}, \bibinfo {author}
  {\bibfnamefont {D.}~\bibnamefont {Vanderbilt}}, \ and\ \bibinfo {author}
  {\bibfnamefont {N.}~\bibnamefont {Marzari}},\ }\href {\doibase
  10.1016/j.cpc.2007.11.016} {\bibfield  {journal} {\bibinfo  {journal}
  {Computer Physics Communications}\ }\textbf {\bibinfo {volume} {178}},\
  \bibinfo {pages} {685} (\bibinfo {year} {2008})}\BibitemShut {NoStop}%
\bibitem [{\citenamefont {Wu}\ \emph {et~al.}(2018)\citenamefont {Wu},
  \citenamefont {Zhang}, \citenamefont {Song}, \citenamefont {Troyer},\ and\
  \citenamefont {Soluyanov}}]{Wu2018}%
  \BibitemOpen
  \bibfield  {author} {\bibinfo {author} {\bibfnamefont {Q.~S.}\ \bibnamefont
  {Wu}}, \bibinfo {author} {\bibfnamefont {S.~N.}\ \bibnamefont {Zhang}},
  \bibinfo {author} {\bibfnamefont {H.~F.}\ \bibnamefont {Song}}, \bibinfo
  {author} {\bibfnamefont {M.}~\bibnamefont {Troyer}}, \ and\ \bibinfo {author}
  {\bibfnamefont {A.~A.}\ \bibnamefont {Soluyanov}},\ }\href {\doibase
  10.1016/j.cpc.2017.09.033} {\bibfield  {journal} {\bibinfo  {journal}
  {Computer Physics Communications}\ }\textbf {\bibinfo {volume} {224}},\
  \bibinfo {pages} {405} (\bibinfo {year} {2018})}\BibitemShut {NoStop}%
\bibitem [{\citenamefont {Xiao}\ \emph {et~al.}(2010)\citenamefont {Xiao},
  \citenamefont {Chang},\ and\ \citenamefont {Niu}}]{Xiao2010}%
  \BibitemOpen
  \bibfield  {author} {\bibinfo {author} {\bibfnamefont {D.}~\bibnamefont
  {Xiao}}, \bibinfo {author} {\bibfnamefont {M.~C.}\ \bibnamefont {Chang}}, \
  and\ \bibinfo {author} {\bibfnamefont {Q.}~\bibnamefont {Niu}},\ }\href
  {\doibase 10.1080/01496396908052235} {\bibfield  {journal} {\bibinfo
  {journal} {Reviews of Modern Physics}\ }\textbf {\bibinfo {volume} {82}},\
  \bibinfo {pages} {1959} (\bibinfo {year} {2010})}\BibitemShut {NoStop}%
\bibitem [{\citenamefont {Tian}\ \emph {et~al.}(2009)\citenamefont {Tian},
  \citenamefont {Ye},\ and\ \citenamefont {Jin}}]{Tian2009}%
  \BibitemOpen
  \bibfield  {author} {\bibinfo {author} {\bibfnamefont {Y.}~\bibnamefont
  {Tian}}, \bibinfo {author} {\bibfnamefont {L.}~\bibnamefont {Ye}}, \ and\
  \bibinfo {author} {\bibfnamefont {X.}~\bibnamefont {Jin}},\ }\href {\doibase
  10.1103/PhysRevLett.103.087206} {\bibfield  {journal} {\bibinfo  {journal}
  {Physical Review Letters}\ }\textbf {\bibinfo {volume} {103}},\ \bibinfo
  {pages} {087206} (\bibinfo {year} {2009})}\BibitemShut {NoStop}%
\bibitem [{\citenamefont {Shitade}\ and\ \citenamefont
  {Nagaosa}(2012)}]{Shitade2012}%
  \BibitemOpen
  \bibfield  {author} {\bibinfo {author} {\bibfnamefont {A.}~\bibnamefont
  {Shitade}}\ and\ \bibinfo {author} {\bibfnamefont {N.}~\bibnamefont
  {Nagaosa}},\ }\href {\doibase 10.1143/JPSJ.81.083704} {\bibfield  {journal}
  {\bibinfo  {journal} {Journal of the Physical Society of Japan}\ }\textbf
  {\bibinfo {volume} {81}},\ \bibinfo {pages} {083704} (\bibinfo {year}
  {2012})}\BibitemShut {NoStop}%
\bibitem [{\citenamefont {{\v{Z}}elezn{\'{y}}}()}]{ZeleznyProgram}%
  \BibitemOpen
  \bibfield  {author} {\bibinfo {author} {\bibfnamefont {J.}~\bibnamefont
  {{\v{Z}}elezn{\'{y}}}},\ }\href {\doibase
  https://bitbucket.org/zeleznyj/linear} {\
  https://bitbucket.org/zeleznyj/linear}\BibitemShut {NoStop}%
\bibitem [{\citenamefont {{\v{S}}mejkal}\ and\ \citenamefont
  {Jungwirth}()}]{Smejkal}%
  \BibitemOpen
  \bibfield  {author} {\bibinfo {author} {\bibfnamefont {L.}~\bibnamefont
  {{\v{S}}mejkal}}\ and\ \bibinfo {author} {\bibfnamefont {T.}~\bibnamefont
  {Jungwirth}},\ }\href {\doibase arXiv:1804.05628 [cond-mat.mes-hall]} {\
  arXiv:1804.05628 [cond-mat.mes-hall]}\BibitemShut {NoStop}%
\bibitem [{\citenamefont {Samathrakis}\ and\ \citenamefont
  {Zhang}()}]{Samathrakis2019}%
  \BibitemOpen
  \bibfield  {author} {\bibinfo {author} {\bibfnamefont {I.}~\bibnamefont
  {Samathrakis}}\ and\ \bibinfo {author} {\bibfnamefont {H.}~\bibnamefont
  {Zhang}},\ }\href@noop {} {\bibinfo  {journal} {(to be published)}\
  }\BibitemShut {NoStop}%
\bibitem [{\citenamefont {Sangiao}\ \emph {et~al.}(2009)\citenamefont
  {Sangiao}, \citenamefont {Morellon}, \citenamefont {Simon}, \citenamefont
  {Teresa}, \citenamefont {Pardo}, \citenamefont {Arbiol},\ and\ \citenamefont
  {Ibarra}}]{Sangiao2009}%
  \BibitemOpen
\bibfield  {journal} {  }\bibfield  {author} {\bibinfo {author} {\bibfnamefont
  {S.}~\bibnamefont {Sangiao}}, \bibinfo {author} {\bibfnamefont
  {L.}~\bibnamefont {Morellon}}, \bibinfo {author} {\bibfnamefont
  {G.}~\bibnamefont {Simon}}, \bibinfo {author} {\bibfnamefont {J.~M.~D.}\
  \bibnamefont {Teresa}}, \bibinfo {author} {\bibfnamefont {J.~A.}\
  \bibnamefont {Pardo}}, \bibinfo {author} {\bibfnamefont {J.}~\bibnamefont
  {Arbiol}}, \ and\ \bibinfo {author} {\bibfnamefont {M.~R.}\ \bibnamefont
  {Ibarra}},\ }\href {\doibase 10.1103/PhysRevB.79.014431} {\bibfield
  {journal} {\bibinfo  {journal} {Physical Review B}\ }\textbf {\bibinfo
  {volume} {79}},\ \bibinfo {pages} {014431} (\bibinfo {year}
  {2009})}\BibitemShut {NoStop}%
\bibitem [{\citenamefont {Wang}\ \emph {et~al.}(2018)\citenamefont {Wang},
  \citenamefont {Xu}, \citenamefont {Lou}, \citenamefont {Liu}, \citenamefont
  {Li}, \citenamefont {Huang}, \citenamefont {Shen}, \citenamefont {Weng},
  \citenamefont {Wang},\ and\ \citenamefont {Lei}}]{Wang2018}%
  \BibitemOpen
  \bibfield  {author} {\bibinfo {author} {\bibfnamefont {Q.}~\bibnamefont
  {Wang}}, \bibinfo {author} {\bibfnamefont {Y.}~\bibnamefont {Xu}}, \bibinfo
  {author} {\bibfnamefont {R.}~\bibnamefont {Lou}}, \bibinfo {author}
  {\bibfnamefont {Z.}~\bibnamefont {Liu}}, \bibinfo {author} {\bibfnamefont
  {M.}~\bibnamefont {Li}}, \bibinfo {author} {\bibfnamefont {Y.}~\bibnamefont
  {Huang}}, \bibinfo {author} {\bibfnamefont {D.}~\bibnamefont {Shen}},
  \bibinfo {author} {\bibfnamefont {H.}~\bibnamefont {Weng}}, \bibinfo {author}
  {\bibfnamefont {S.}~\bibnamefont {Wang}}, \ and\ \bibinfo {author}
  {\bibfnamefont {H.}~\bibnamefont {Lei}},\ }\href {\doibase
  10.1038/s41467-018-06088-2} {\bibfield  {journal} {\bibinfo  {journal}
  {Nature Communications}\ }\textbf {\bibinfo {volume} {9}},\ \bibinfo {pages}
  {3681} (\bibinfo {year} {2018})}\BibitemShut {NoStop}%
\bibitem [{\citenamefont {Hal{\'{a}}sz}\ and\ \citenamefont
  {Balents}(2012)}]{Halasz2012}%
  \BibitemOpen
  \bibfield  {author} {\bibinfo {author} {\bibfnamefont {G.~B.}\ \bibnamefont
  {Hal{\'{a}}sz}}\ and\ \bibinfo {author} {\bibfnamefont {L.}~\bibnamefont
  {Balents}},\ }\href {\doibase 10.1103/PhysRevB.85.035103} {\bibfield
  {journal} {\bibinfo  {journal} {Physical Review B}\ }\textbf {\bibinfo
  {volume} {85}},\ \bibinfo {pages} {035103} (\bibinfo {year}
  {2012})}\BibitemShut {NoStop}%
\bibitem [{\citenamefont {Kida}\ \emph {et~al.}(2011)\citenamefont {Kida},
  \citenamefont {Fenner}, \citenamefont {Dee}, \citenamefont {Terasaki},
  \citenamefont {Hagiwara},\ and\ \citenamefont {Wills}}]{Kida2011}%
  \BibitemOpen
  \bibfield  {author} {\bibinfo {author} {\bibfnamefont {T.}~\bibnamefont
  {Kida}}, \bibinfo {author} {\bibfnamefont {L.~A.}\ \bibnamefont {Fenner}},
  \bibinfo {author} {\bibfnamefont {A.~A.}\ \bibnamefont {Dee}}, \bibinfo
  {author} {\bibfnamefont {I.}~\bibnamefont {Terasaki}}, \bibinfo {author}
  {\bibfnamefont {M.}~\bibnamefont {Hagiwara}}, \ and\ \bibinfo {author}
  {\bibfnamefont {A.~S.}\ \bibnamefont {Wills}},\ }\href {\doibase
  10.1088/0953-8984/23/11/112205} {\bibfield  {journal} {\bibinfo  {journal}
  {Journal of Physics Condensed Matter}\ }\textbf {\bibinfo {volume} {23}},\
  \bibinfo {pages} {112205} (\bibinfo {year} {2011})}\BibitemShut {NoStop}%
\end{thebibliography}

%

\newpage
\newcommand{\beginsupplement}{%
        \setcounter{table}{0}
        \renewcommand{\thetable}{S\arabic{table}}%
        \setcounter{figure}{0}
        \renewcommand{\thefigure}{S\arabic{figure}}%
     }
\beginsupplement


\begin{figure}
\includegraphics[width=0.4\textwidth]{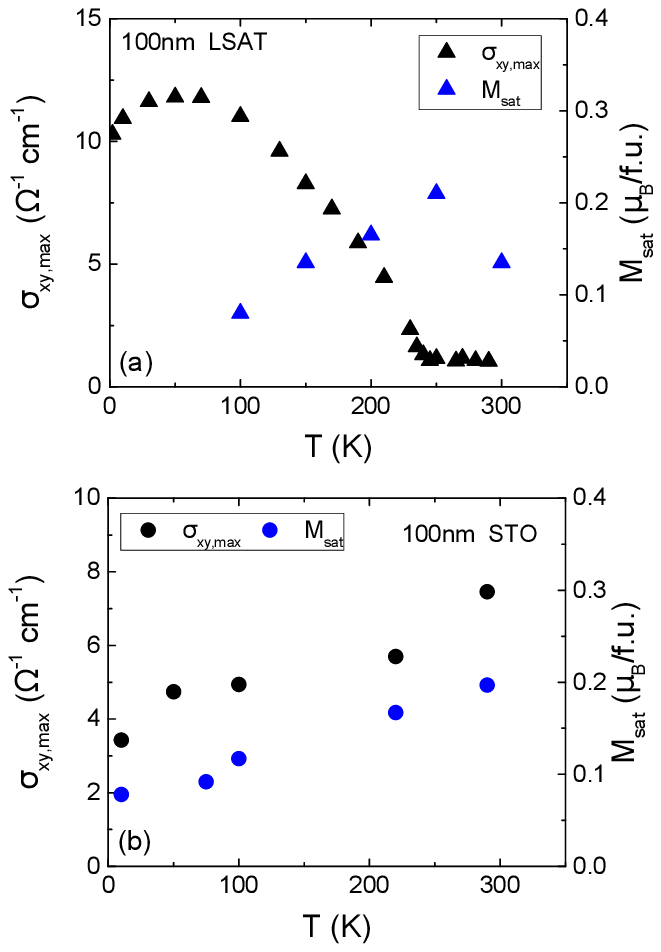}%
\caption{Temperature dependence of $\sigma_{xy,\mathrm{max}}$ and $M_{\mathrm{sat}}$ for 100\,nm Mn$_{3}$NiN films grown on (a) LSAT and (b) STO.}
\label{FigS1}
\end{figure}

\begin{figure}
\includegraphics[width=0.4\textwidth]{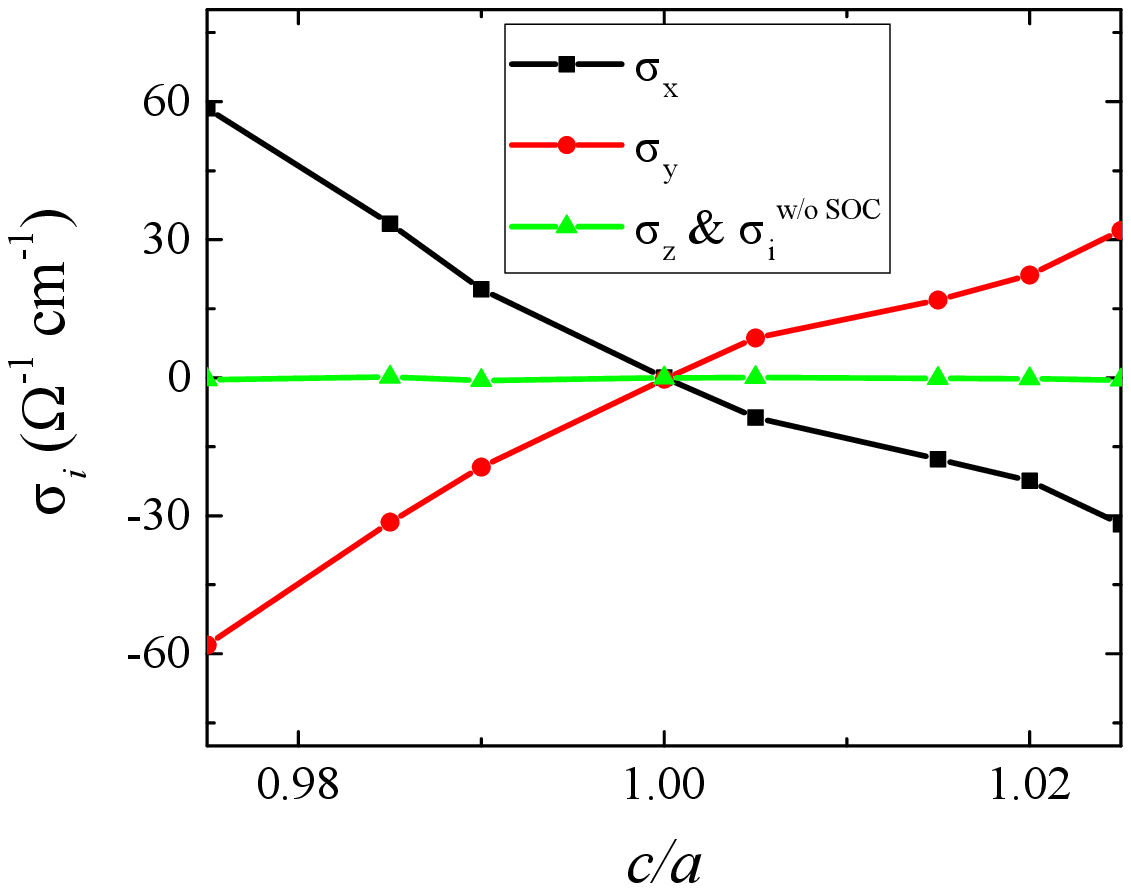}%
\caption{The behaviour of the calculated intrinsic anomalous Hall conductivity components, $\sigma_{i}$, under strain for the \Gfive\ structure.}
\label{FigS2}
\end{figure}

\end{document}